\documentclass[aps,prl,preprint,superscriptaddress]{revtex4}
\usepackage{graphicx}

\begin{document}
\title{Quantum capacitance anomalies of two-dimensional non-equilibrium states under microwave irradiation}
\author{Jian Mi}
\affiliation{International Center for Quantum Materials, Peking University, Beijing, 100871, China}

\author{Jianli Wang}
\affiliation{International Center for Quantum Materials, Peking University, Beijing, 100871, China}

\author{Saeed Fallahi}
\affiliation{Department of Physics and Astronomy, and Station Q Purdue, Purdue University, West Lafayette, Indiana 47907, USA}
\affiliation{Birck Nanotechnology Center, Purdue University, West Lafayette, Indiana 47907, USA}

\author{Geoffrey C. Gardner}
\affiliation{Department of Physics and Astronomy, and Station Q Purdue, Purdue University, West Lafayette, Indiana 47907, USA}
\affiliation{Birck Nanotechnology Center, Purdue University, West Lafayette, Indiana 47907, USA}
\affiliation{School of Electrical and Computer Engineering, and School of Materials Engineering, Purdue University, Indiana 47907, USA}

\author{M. J. Manfra}
\affiliation{Department of Physics and Astronomy, and Station Q Purdue, Purdue University, West Lafayette, Indiana 47907, USA}
\affiliation{Birck Nanotechnology Center, Purdue University, West Lafayette, Indiana 47907, USA}
\affiliation{School of Electrical and Computer Engineering, and School of Materials Engineering, Purdue University, Indiana 47907, USA}

\author{Chi Zhang}
\altaffiliation{Electronic address: gwlzhangchi@pku.edu.cn}
\affiliation{International Center for Quantum Materials, Peking University, Beijing, 100871, China}
\affiliation{Collaborative Innovation Center of Quantum Matter, Beijing, 100871, China}



\begin{abstract}
  We report our direct study of the compressibility on ultrahigh mobility two-dimensional electron system ($\mu_{e} \sim$ $1 \times 10^7$ cm$^2$/Vs) in GaAs/AlGaAs quantum wells under microwave (MW) irradiation.
  The field penetration current results show that the quantum capacitance oscillates with microwave induced resistance oscillations (MIRO), however, the trend is opposite with respect to the compressibility for usual equilibrium states in previous theoretical explanations.
  The anomalous phenomena provide a platform for study on the non-equilibrium system under microwave, and point to the current domains and inhomogeneity induced by radiation.
  Moreover, the quantum capacitance indication for multi-photon process around $j = 1/2$ is detected under intensive microwave below 30 GHz.
\end{abstract}

\maketitle

Two-dimensional electron system (2DES) has always been one of the important platforms in condensed matter physics research in the past decades.
In particular, microwave-induced resistance oscillations (MIRO) ~\cite{Zudov2001, YeAPL2001} and zero-resistance states (ZRS) ~\cite{Mani,Zudov} were discovered, for ultrahigh mobility 2DES samples under microwave (MW) irradiation and low magnetic field ($B$).
The oscillations can be expressed as a function of $j \equiv \omega / \omega_{c}$, where $\omega$ is the microwave frequency, $\omega_{c} = eB/m^{*}$ is the cyclotron frequency, and $m^{*}$ is the effective mass of electron.
Much attention has been paid in search of the physical origin of MIRO, and the most popular theoretical models include the displacement model ~\cite{Durst2003}, the inelastic model ~\cite{Dimitriev2005} and the radiation driven electron orbit model ~\cite{Inarrea}.
Theoretically, the vanishing resistance emerges from unstable negative resistance, and current domains are formed under microwave ~\cite{AndreevPRL2003}.
A time-dependent photovoltage experiment has implied the existence of domain walls ~\cite{DorozhkinNP2011}, but no direct study has provided confirmative evidence for the current domains.

Most experiments about MIRO and ZRS are focused on magnetoresistances or magnetoconductances.
The limitation of this method lies in the contacts and the conducting current effects on the non-equilibrium state ~\cite{Mikhailov2011}.
Other reported methods proposed thermoelectric measurements ~\cite{Levin2015} and capacitance measurements ~\cite{Levin2016}.

In 2DES, quantum capacitance is proportional to the density of states (DOS) and the compressibility ($\kappa$): $C_{q} = e^2(dn/d\mu) = e^2 n^2 \kappa$ ~\cite{Luryi, JPE1992}, where $n$ is the carrier density and $\mu$ is the chemical potential.
A theoretical study on MIRO indicates that the DOS oscillates with the longitudinal resistance and becomes incompressible in ZRS ~\cite{Vavilov2004}.
Experiments based on a single electron transistor (SET) suggest the oscillatory properties of local compressibility ~\cite{Nuebler}, but the results contain irregularities and vary with the SET position and microwave frequency.
In this study, we measure the quantum capacitance of high mobility 2DES under MW irradiation.

The prototype method of studying quantum capacitance $C_{q}$ is to directly measure the capacitance $C$ between 2DES and a gate electrode, $\frac{1}{C}=\frac{1}{C_{q}}+\frac{1}{C_{g}}$ ($C_{g}$ is the geometric capacitance).
But it is difficult to extract $C_{q}$ from $C$, because $C_{q} \gg C_{g}$.
A modified technique is the electric field penetration method ~\cite{JPE1992}, which has been successfully applied to study the interacting two-dimensional systems ~\cite{Dultz-Jiang, LuLi}.
We adopt this method to measure the quantum capacitance of 2DES under MW irradiation.
Our results show the oscillating features of $C_{q}$, but $C_{q}$ exhibits opposite trends with the theory ~\cite{Vavilov2004}.
The penetrating current shows a minimum at ZRS, and the intensity of current decreases under a high power microwave.
These observations point to the existence of microwave-induced current domains and inhomogeneity.

The wafer used in our experiments is a high-quality, \emph{in-situ} back-gated GaAs/Al$_{0.24}$Ga$_{0.76}$As quantum well (QW) grown by molecular-beam epitaxy ~\cite{Watson-Manfra}.
The 30 nm wide QW is located about 200 nm beneath the sample surface, and the highly n$^{+}$-doped \emph{in-situ} gate is 850 nm below the QW.
The electron density and mobility are $n_{e} \sim 1.6 \times 10^{11}$ cm$^{-2}$ and $\mu_{e} \sim 1.0 \times 10^{7}$ cm$^{2}$/Vs respectively below 4 K.
In 2D carriers, the Coulomb energy exceeds the kinetic energy at low density, and the dimensionless parameter $r_{s}=\sqrt{\pi n}/ a_{B}$ describes the ratio between the two.
The ratio $r_{s}$ is 1.4, with an electron Bohr radius ($a_{B}$) of 10.2 nm in our GaAs device.

To fabricate a device for our study, we define an 800 $\mu$m $\times$ 800 $\mu$m square mesa with 6 arms by UV-lithography and wet etching.
Ohmic contacts are made by an 8/80/160/36 nm stack of Ni/Ge/Au/Ni metals.
The alloy front-gate is deposited on the device which is transparent for MW irradiation.
The measurements are carried out in a He3 refrigerator with a base temperature of 0.3 K, and the microwave is guided down to the base via a WR-28 waveguide.
In the $C_{q}$ measurements, the 2DES is grounded to screen the electric field.
We apply a 20 mV ac-excitation $V_{ac}$ to the back-gate, and detect the penetrating current $I_{p}$ from the front-gate.
The experimental setup with the sandwich-like device structure is illustrated in the inset of FIG. 1.
According to Dultz and Jiang ~\cite{Dultz-Jiang}, in the low frequency limit, the in-phase component $I_{x}$ is proportional to $-1/\sigma_{xx}$ ($\sigma_{xx}$: conductivity), and the 90$^{o}$ phase $I_{y}$ ($I_{y} \equiv I_{p}$) is proportional to $1/C_{q}$ or 1/$\kappa$.

\begin{figure}
\includegraphics[width=0.8\linewidth]{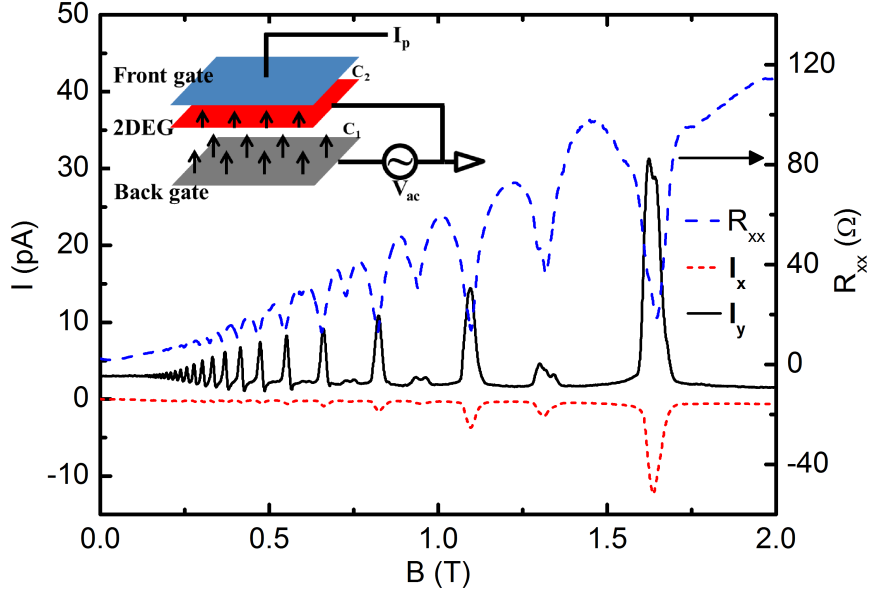}
\caption{(Color online). Magnetoresistance, $x$-component $I_{x}$ and (90$^{o}$ phase) $y$-component $I_{y}$ of the penetration current at 300 mK. The inset shows the sketch of the measurement setup.}
  \label{FIG1}
 \end{figure}

We comparatively study the quantum capacitances for the equilibrium states: the Shubnikov-de Haas (SdH) oscillations and the integer quantum Hall effects (IQHE).
Figure 1 shows the traces of $I_{x}$, $I_{y}$, and magnetoresistance $R_{xx}$ as a function of $B$ (without irradiation) at 300 mK.
The frequency of ac-voltage $V_{ac}$ is 91 Hz.
The $R_{xx}$ exhibits IQHE features and reaches minima at the integer fillings.
Meanwhile, the current components $I_{x}$ and $I_{y}$ show maxima of their absolute values, indicating that $\kappa$ and $\sigma_{xx}$ are approaching zero in the IQHE insulating regime.
When the Fermi level is located in the gap between two adjacent Landau levels, the DOS tends to be zero, and the electron states become more incompressible ($\kappa \downarrow$).
These experimental results can be easily explained with the classical theory of Landau levels.
When the Fermi level is located in the center of a Landau level, $I_{y}$ presents a minimum, and the compressibility reaches a maximum.

The method of the quantum capacitance study can be simplified ~\cite{Luryi, Dultz-Jiang}, as is shown in the inset of FIG. 1, where $C_{q}$ is illustrated with the geometric capacitances $C_{1}$ and $C_{2}$.
Based on the $I$-$V$ characterization, we estimate the capacitances of $C_{1} = 360$ pF and $C_{2} = 45$ pF; and a reasonable estimation of $C_{q}$ is 63 nF in the absence of magnetic fields.
So far we have provided the proof for the validity and quality of our devices; $I_{x}$ and $I_{y}$ both meet the ac-frequency criteria of the quantum capacitance model ~\cite{Dultz-Jiang}, (the details are shown in the Supplementary Information-SI).

In the regime of SdH oscillations and IQHE, the experimental data show distinct features of $I_{x}$ and $I_{y}$ components.
We extend the field penetration method to the non-equilibrium states.
Figure 2(A) presents the longitudinal resistance under a 40 GHz MW at 0.5 K.
$R_{xx}$ shows strong MIRO features and a tendency to form ZRS.
Figure 2(A) also shows the penetrating currents $I_{y}$ with and without MW irradiation (the green and the grey curves).
We notice that $I_{y}$ oscillates weakly with magnetoresistance when the device is irradiated by MW.
However, sharply different from the SdH oscillations, the current $I_{y}$ exhibits opposite trends.
It reaches maximal values at the maximum resistances of MIRO.
On the contrary, at the maximal resistances of SdH oscillations, $I_{y}$ reaches minima, which is consistent with the Landau level spectrum.
In the low $B$ MIRO regime at 40 GHz, $C_{q}$ decreases at the resistance maximum, and increases at the $R_{xx}$ minimum.
Phenomenologically, a theoretical work proposed that the electron system becomes incompressible ($\kappa \downarrow$) at the resistance minimum ~\cite{Vavilov2004}.
Considering that the $I_{p}$ signals are very small and noisy at low magnetic fields, we use the Fast Fourier Transform (FFT) method to filter out noises in order to highlight the MIRO signals.
The averaged results for the sample with and without MW irradiation are displayed by the red and the black curves in FIG. 2(A).
It is evident that the averaged trace (envelope) in the MIRO regime is not weaker than the amplitudes in the SdH oscillations regime.
The change of the penetrating current $\Delta I_{p}$ at 40 GHz is illustrated in Panel (B), whose background is subtracted.
The $\Delta I_{p}$ maxima are distinct around the integer $j \sim 1, 2$, and the $\Delta I_{p}$ minima exist around $j \sim 5/4$ (marked by arrows).

\begin{figure}
\includegraphics[width=0.8\linewidth]{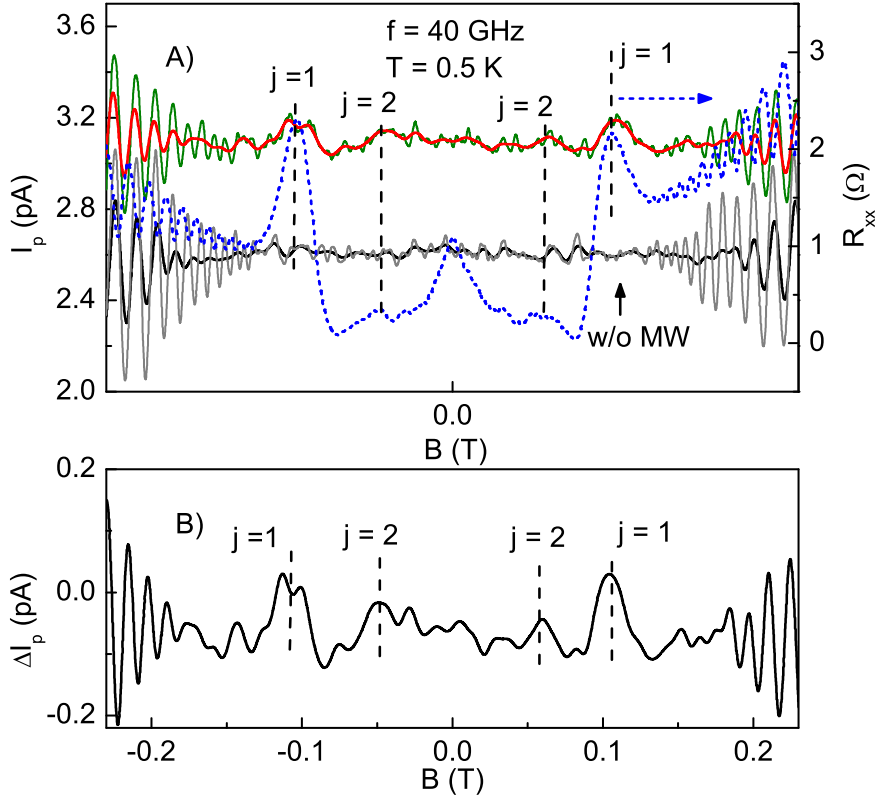}
\caption{(Color online). (A): The magnetoresistance (blue dotted curve) and the penetrating current $I_{p}$ under a 40 GHz MW. The penetrating current without MW is also shown (by the grey color curve). The red and the black $I_{p}-$curves indicate the smoothed results by the FFT method. (B): The trace for the change of penetrating $\Delta I_{p}$ versus magnetic field.}
  \label{FIG2}
 \end{figure}

\begin{figure}
\includegraphics[width=0.8\linewidth]{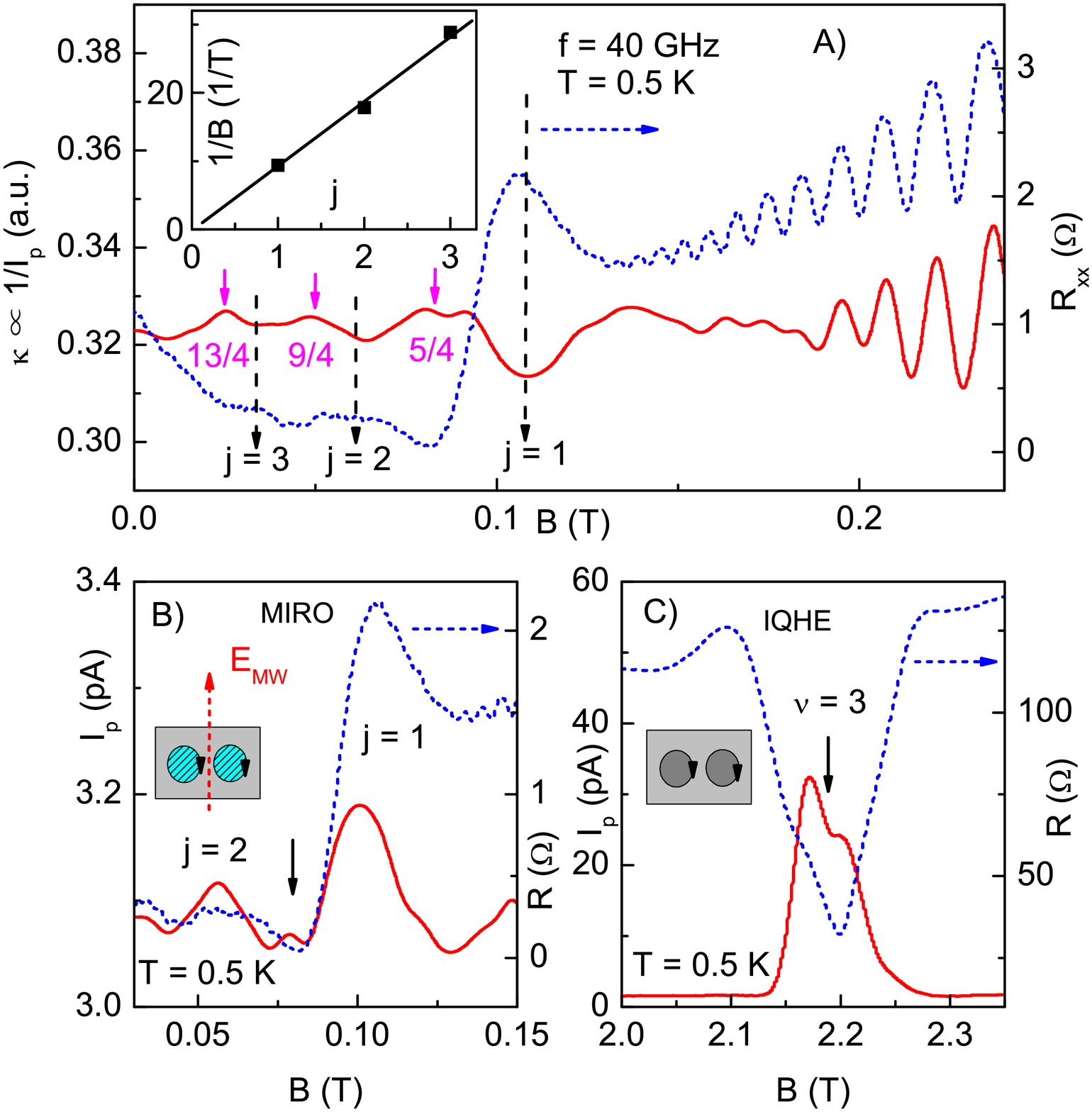}
\caption{(Color online). (A): The compressibility ($\kappa \propto 1/I_{p}$) and the magnetoresistane (blue dotted curve) under 40 GHz microwave: the minima exist around the integer $j \sim 1, 2, 3$. The relation of $1/B$ versus integer $j$ is shown in the inset. (B): The penetrating current and MR between $j = 1$ and $2$. The arrow marks ZRS around $j \sim 5/4$. The 2DES sample under magnetic field and electric field of microwave is shown in the inset. (C): $I_{p}$ and $R_{xx}$ for the IQHE regime $\nu \sim 3$ at 0.5 K. The sample under $B$-field with MW power off is shown in the inset.}
  \label{FIG3}
 \end{figure}

The specific features of MW-induced non-equilibrium are more visible if we comparatively study $\kappa$ ($\propto 1/I_{y}$) and $R_{xx}$ as shown in FIG. 3(A).
The plot can be separated into two parts: the MIRO regime and the SdH oscillation regime.
The MIRO features dominate below 0.13 T, and the SdH oscillations are fully developed above 0.13 T.
Qualitatively, the $\kappa$ minima appear around maximal resistances in the MIRO regime, whereas the maximal DOS exists at maximal MR in the SdH regime.
Quantitatively, the amplitudes of $1/C_{q}$ oscillations ($j = 1, 2, 3$) are not weaker than those of SdH oscillations.
The $j = 1, 2$ amplitudes of $\kappa$ are quite large, and the weaker $j = 3$ amplitude is as large as those at the onset of SdH oscillations.
The maximal $\kappa$ around $j \sim 5/4, 9/4, 13/4$ are marked by the pink color solid arrows.

In light of the similarity between MIRO and IQHE (or SdH), a comparative study on field penetration in these two effects is in order.
The MR and the penetrating current for MIRO and SdH are shown in FIG. 3(B) and (C) respectively.
The magnetoresistances for MIRO with $j \sim (1 - 2)$ in panel (B) resemble those for IQHE around integer fillings ($\nu \sim 3$) in panel (C).
The MIRO $I_{p}$-trace has a ``W"-shape at the regime of ZRS between $j = 1$ and 2.
In contrast, the gap state at $\nu \sim 3$ exhibits an ``M"-shape current feature.
The current minima and maxima are marked by arrows in both panels, which also highlight the slope turning points for $R_{xx}$.
It seems that the measured currents for MIRO-ZRS form an opposite pattern from those for IQHE.
The MW-induced current domains (green color) and the localized electron in the bulk state are shown in the insets of Panel (B) and (C), respectively.
The light grey color rectangle represents the 2DES plane, to which the applied $B$ is perpendicular.
The microwave electric-vector $E_{MW}$ in the waveguide is illustrated in the inset of Panel (B) ~\cite{Zudov2001}.
Due to the existence of a net current $j_{0}$, a conducting state exists within each current domain ~\cite{AndreevPRL2003}.
In usual transport for IQHE, the electrons are localized in the bulk (inset of FIG. 3(C)).

In our device, the microwave is weakened by the metallic top-gate to some degree.
To compensate for this, the quantum capacitance signals can be strengthened by raising the MW power ($P$).
The $P$-dependent penetrating currents at 30 GHz MW are shown in FIG. 4(A).
At first sight, the relation between magnetoresistance and quantum capacitance for 30 GHz microwave is very similar to that for 40 GHz shown in FIG. 2.
The amplitudes of SdH oscillations are weakened by MW power owing to the heating effect.
The peak position of $\Delta I_{p}$ shifts slightly to higher $B$ and the minimum position shifts slightly to lower $B$ with power, which is in agreement with the $P$-dependent MR features ~\cite{Zudov} and can be explained by the enhancement and broadening of resonance peak in high MW power.
The $C_{q}$ oscillations can be strengthened gradually by increasing MW power.
However, at the highest power regime above 15 dBm, the $C_{q}$ oscillations can be weakened by the heating effect on electrons.

Here we guide the eyes by a pair of (dotted) reference lines for each $\Delta I_{p}$ trace in FIG. 4(A).
The top line is aligned with the $j = 1$ maximum, and the bottom one marks the midpoint of the SdH oscillations.
Under microwave, the SdH amplitudes are weakened by a factor of $(20 \sim 35)\%$.
The $\Delta I_{p}$ intensities at $j = 1$ are almost the same for different powers.
The minima at $j \sim 5/4$ decrease as the microwave power with respect to the $j = 1$ maxima or the top guidelines.
The light yellow color stripes are to highlight that the minimal $\Delta I_{p}$ regime is wider under higher power microwave.
Based on the theoretical description ~\cite{AndreevPRL2003} and the experimental study ~\cite{DorozhkinNP2011} on the ZRS, the current domain is expected to be robust under high microwave power.
Our observations for the width of minimal penetrating current at ZRS are consistent with the power-dependent telegraph signals ~\cite{DorozhkinNP2011}.

In addition, in the traces of 8 dBm and 12 dBm, we observe the indication of multi-photon processes at $j = 1/2$ ~\cite{Zudov2006}, which is marked by red arrows in FIG. 4(A).
This feature appears only under high MW power, it cannot be detected in magnetoresistances in our gated devices.
As far as we know, the multi-photon process for the MIRO has only been reported below 30 GHz.
In transport, the current domains and patterns are strongly weakened by the external dc-current ~\cite{DorozhkinNP2011}.
Therefore we obtain more current domains information in quantum capacitance than in transport measurements.

\begin{figure}
\includegraphics[width=0.8\linewidth]{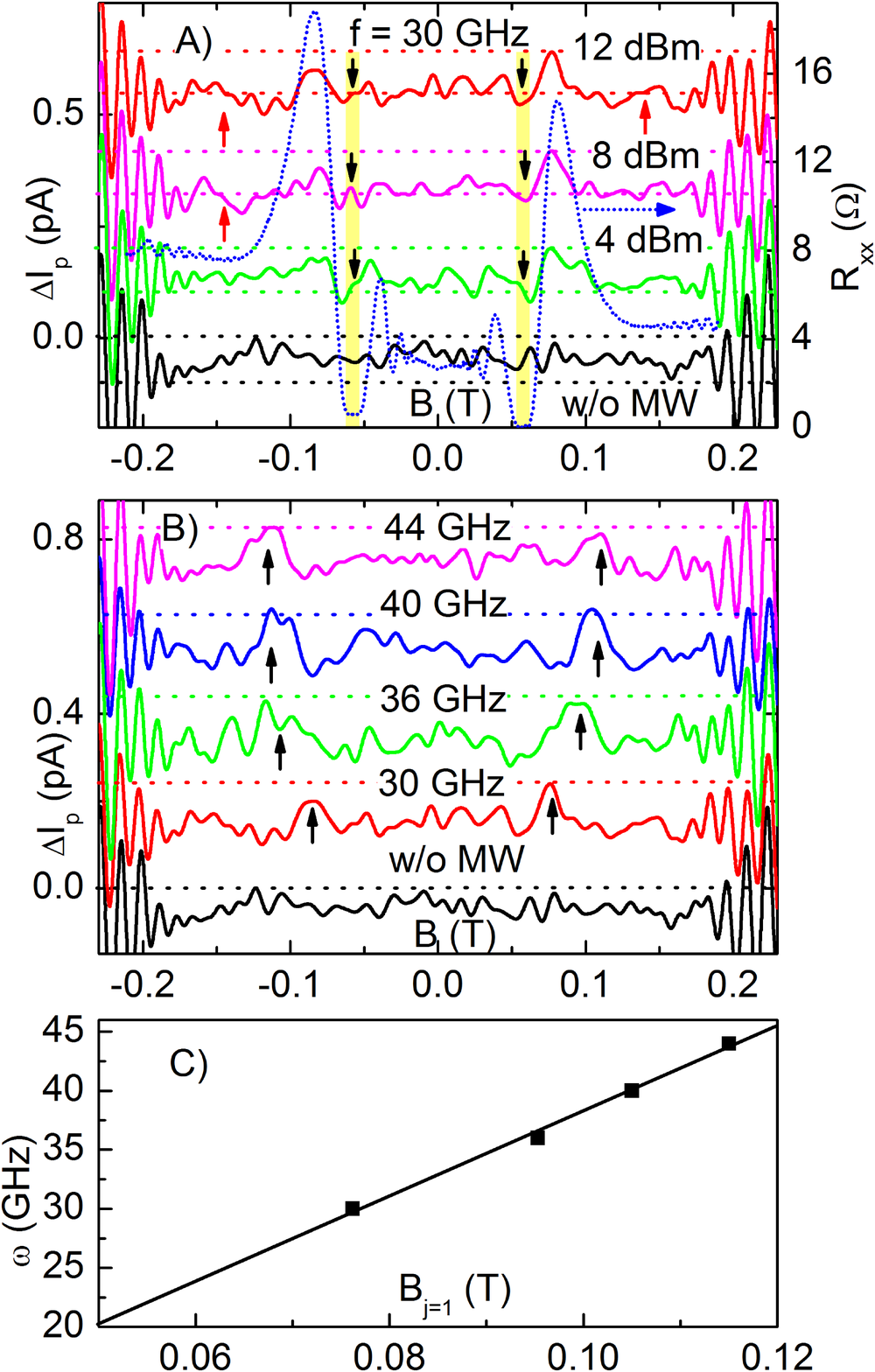}
\caption{(Color online). (A) Power-dependent $\Delta I_{p}$ under 30 GHz MW: MW-off, 4, 8, 12 dBm curves are in black, green, pink, red color respectively. The black downward arrows mark the ZRS, and the red arrows mark the multi-photon process at $j = 1/2$ at high power. $R_{xx}$ is shown by the blue dotted curve. (B) The frequency-dependent $\Delta I_{p}$ at 30 - 44 GHz. The indications at $j=1$ are marked by arrows. (C) The MW frequency vs. $B$-field for $j = 1$ is shown.}
  \label{FIG4}
\end{figure}

The penetrating currents for MIRO and ZRS at $f \sim 30 - 44$ GHz are presented in FIG. 4(B).
The $\Delta I_{p}$ traces are very symmetric with respect to magnetic fields.
Each microwave-$f$ curve exhibits a peak at $j = 1$ and a valley in the region of $j \sim 1 - 2$.
With increasing MW frequency, the $\Delta I_{p}$ expands along the $B$-axis.
In addition, as shown in Panel (C), the $j = 1$ peak position in magnetic field is proportional to MW frequency, which is in accordance with magnetoresistance results quantitatively.
Similar findings are obtained even for much higher frequency ($\sim 102$ GHz, see SI).
Although the guide of high frequency power is less efficient, indications of the penetrating current are still clearly observed, especially for the minima around $j \sim 5/4$.

The compressibility for microwave induced non-equilibrium electron states has never been directly measured in experiments.
Recent development on very dilute (non-degenerate) electrons on Helium indicates that the transitions between the two sets of shifting Landau level are incompressible ~\cite{Chepeliankii}; another current preprint shows a similar case, which reports a microwave beating pattern from the second electron subband ~\cite{Dorozhkin2016}.
However, the compressibility of direct photon transition for the MIRO/ZRS remains unstudied.
In magnetoresistances, the maxima appear close to the integer $j = n$, and the minima are near $j = n + 1/4$ ~\cite{ZudovRMP}.
In quantum capacitances, the $\Delta I_{p}$ spike at $j \sim 1$ is largely due to the distinct relative decrease of current nearby $j \sim 5/4$ ZRS regime.
Under an intensive MW, the charge inhomogeneities are inevitable, and the current flow in the form of domain patterns ~\cite{DorozhkinNP2011}.
Each domain carries a dissipationless current density $j_{0}$, and the dissipative electric field is quenched ~\cite{AndreevPRL2003}.
The conducting domains in the electron channel lead to the decreasing $\Delta I_{p}$ and the increasing DOS around minimal resistances.

The explanation for this observed quantum capacitance anomalies at oscillations/ZRS should be focused on the properties of non-equilibrium state.
In the theory ~\cite{AndreevPRL2003}, the current domains induced by microwave are metallic, and the net current accumulation at the current domains leads to the charge inhomogeneity.
According to Efros ~\cite{Efros2008}, in a strongly non-ideal electron system ($r_{s} \geq 1$) with a long-range interaction, the inhomogeneity is a source of the negative trend in compressibility signals.
The electric force from inhomogeneous density causes density relaxation much faster than diffusion, which leads to the negative or opposite trends of compressibility.

Our study has performed direct experiments on the quantum capacitance (the compressibility) of the microwave induced non-equilibrium states in an ultraclean 2DEG.
Qualitatively, the non-equilibrium states compressibilities show opposite trends with respect to those of the SdH oscillations and IQHE.
And the same puzzling phenomenon can be reproduced systematically in different samples and devices.
Our project may provide direct evidence for the existence of microwave-driven charge inhomogeneity or current domains.
Moreover, the quantum capacitance provides another perspective to the 2D non-equilibrium states: the multiphoton process is observable at $j = 1/2$, while cannot be directly observed in the magnetoresistance.

This project at Peking University is supported by National Basic Research Program of China (Grant Nos. 2013CB921903 and 2014CB920904) and the National Science Foundation of China (Grant No.11374020). The work at Purdue was supported by the U. S. Department of Energy, Office of Science, Basic Energy Sciences, under Award No. DE-SC0006671. Additional support from the W. M. Keck foundation is also gratefully acknowledged.

\end{document}